\documentclass[prx,letterpaper,nofootinbib,superscriptaddress,aps,floatfix,twocolumn]{revtex4-1}
\usepackage{graphicx}
\usepackage{amsmath}
\usepackage{amsfonts}
\usepackage{color, colortbl}
\usepackage[small,bf]{subfigure}
\usepackage{cancel}
\usepackage{soul}
\newcommand{\beq}{\begin{equation}}
\newcommand{\eeq}{\end{equation}}
\newcommand{\bk}{{{\bf{k}}}}

\newcommand{\beqa}{\begin{eqnarray}}
\newcommand{\eeqa}{\end{eqnarray}}
\newcommand{\pdg}{{\vphantom \dag}}
\newcommand{\dg}{{\dag}}

\newcommand{\ra}{\rightarrow}

\newcommand{\A}{\mathcal{A}}
\begin{document}
\title{Emergent anomalies and generalized Luttinger theorems in metals and semimetals}
\author{Chong Wang}
\affiliation{Perimeter Institute for Theoretical Physics, Waterloo, Ontario N2L 2Y5, Canada}
\author{Alexander Hickey}
\affiliation{Department of Physics and Astronomy, University of Waterloo, Waterloo, Ontario 
N2L 3G1, Canada} 
\author{Xuzhe Ying}
\affiliation{Department of Physics and Astronomy, University of Waterloo, Waterloo, Ontario 
N2L 3G1, Canada} 
\author{A.A. Burkov}
\affiliation{Department of Physics and Astronomy, University of Waterloo, Waterloo, Ontario 
N2L 3G1, Canada} 
\affiliation{Perimeter Institute for Theoretical Physics, Waterloo, Ontario N2L 2Y5, Canada}
\date{\today} 
\begin{abstract}
Luttinger's theorem connects a basic microscopic property of a given metallic crystalline material, the number of electrons per unit cell, 
to the volume, enclosed by its Fermi surface, which defines its low-energy observable properties.
Such statements are valuable since, in general, deducing a low-energy description from microscopics, which may perhaps be regarded as the main 
problem of condensed matter theory, is far from easy.
In this paper we present a unified framework, which allows one to discuss Luttinger theorems for ordinary metals, as well as closely analogous 
exact statements for topological (semi)metals, whose low-energy description contains either discrete point or continuous line nodes.
This framework is based on the 't Hooft anomaly of the emergent charge conservation symmetry at each point on the Fermi surface, a concept recently proposed by Else, Thorngren and Senthil [Phys. Rev. X {\bf 11}, 021005 (2021)]. We find that the Fermi surface codimension $p$ plays a crucial role for the emergent anomaly. For odd $p$, such as ordinary metals ($p=1$) and magnetic Weyl semimetals ($p=3$), the emergent symmetry has a generalized chiral anomaly. For even $p$, such as graphene and nodal line semimetals (both with $p=2$), the emergent symmetry has a generalized parity anomaly. When restricted to microscopic symmetries, such as $U(1)$ and lattice symmetries, the emergent anomalies imply (generalized) Luttinger theorems, relating Fermi surface volume to various topological responses. The corresponding topological responses are the charge density for $p=1$, Hall conductivity for $p=3$, and polarization for $p=2$. As a by-product of our results, we clarify exactly what is anomalous about the surface states of nodal line semimetals. 
\end{abstract}
\maketitle
\section{Introduction}
\label{sec:1}
The metal-insulator dichotomy is one of the basic facts of solid state physics. 
It is also one of the simplest and most fundamental macroscopic manifestations of quantum mechanics. 
If electrons are treated as noninteracting, the existence of metals and insulators follows from the quantum mechanics 
of a particle in periodic potential, which leads to the existence of bands and band gaps in the electron energy spectrum, 
and from the Pauli principle. 
Whether a given crystalline material is a metal or an insulator is then determined by a single parameter, the number of electrons per unit cell per spin (filling). 
When the number is an integer, the material is an insulator (we will not distinguish between insulators and accidental compensated semimetals due to band overlap). Otherwise it is a metal, with a Fermi surface of gapless excitations, which encloses a volume in momentum space, directly determined by the 
fractional part of the filling, which we denote as $\nu$ henceforth. 

The relation between the Fermi surface volume and the filling is in fact much deeper than simple counting of filled single-particle states of noninteracting 
electrons might suggest. It holds even when the electron-electron interactions are taken into account, a statement known as Luttinger's 
theorem~\cite{Luttinger60}.
Recent work has made it clear that this statement has a topological origin~\cite{Volovik03,Volovik07,Oshikawa00,Hastings04,Furuya17,Cheng16,Cho17,Metlitski18,Jian18,Song21,Else21}, in particular connecting it with the 
concepts of quantum anomalies and higher-dimensional symmetry protected topological (SPT) phases. 
In this sense $\nu$ may be viewed as a topological, although unquantized and continuously tunable, invariant, characterizing a metallic phase, 
which is sandwiched between two insulators, corresponding to integer values of $\nu$. 

On another front, it has recently been discovered that there exists another way a metallic phase may arise: as an intermediate phase between a topological 
and an ordinary insulator in three dimensions (3D), when a direct transition is impossible~\cite{Volovik03,Volovik07,Murakami07,Wan11,Burkov11-1,Burkov11-2,Weyl_RMP}.
Such materials are known as topological semimetals, or topological metals (``metal" in this paper refers to a state with gapless fermionic excitations, but 
not necessarily compressible), and exist at integer electron fillings per unit cell, which would normally 
imply an insulator as the Luttinger invariant $\nu$ is zero. In this case however, there exist other topological invariants, analogous to $\nu$~\cite{Gioia21}.
Just like $\nu$, these invariants are unquantized and continuously tunable, but are topological in origin and require gapless modes to be present. 
In the simplest case of a magnetic Weyl semimetal, the unquantized invariant is the Hall conductance per atomic plane~\cite{Burkov11-1},
which, when not equal to an integer multiple of $e^2/h$, requires gapless Weyl points whenever the Luttinger invariant $\nu$ is zero. 
In other types of topological semimetals the invariants are more subtle, taking the form of fractional electric charges carried by topological 
defects of crystalline symmetries in point node semimetals~\cite{Gioia21}, or fractional polarization in the line node case~\cite{Hughes_linenode,Schnyder18}.
In all cases, however, the idea is very similar to the Luttinger invariant in ordinary metals: a ``fractional" value of the invariant necessarily requires 
either gapless modes or topological order (i.e. long-range entanglement). 
The invariant is continuously tunable between two ``trivial" values, which correspond to insulators. 
These are either ordinary insulators at different integer unit cell filling in the case of ordinary metals, or insulators with different electronic structure topology in the case of topological semimetals. 

This similarity suggests that there should exist a common framework to describe the topological properties of all metallic phases, including both 
ordinary and topological metals. The goal of this paper is to describe such a framework. 
We demonstrate that metals may be described by 't Hooft anomalies of emergent symmetry groups, which characterize their low-energy 
excitation spectrum. 
The 't Hooft anomalies connect the corresponding metallic phases to higher-dimensional 
SPT insulators. 
While the SPT insulators themselves are characterized by quantized topological response terms, the unquantized anomalies in metals 
arise due to unquantized charges of gauged crystalline symmetries (such as translations and rotations), which enter the corresponding 
quantized response terms. 

The nature of the emergent anomaly and the corresponding Luttinger-like topological invariants is decided by the codimension of the Fermi surface $p=d-d_{FS}$, where $d$ is the space dimension and $d_{FS}$ is the Fermi surface dimension. For odd $p$, the emergent symmetry has a generalized chiral anomaly --- examples include ordinary metals ($p=1$) and magnetic Weyl semimetal in 3D ($p=3$). The corresponding topological invariants in the (generalized) Luttinger theorems are the charge density ($p=1$) and the Hall conductivity ($p=3$) respectively. For even $p$, the emergent symmetry has a generalized parity anomaly. Examples with $p=2$ include nodal (Dirac) semimetals in two dimensions (2D) and nodal line semimetals in 3D. In both cases, we show that the corresponding topological invariant in the generalized Luttinger theorem is the electric polarization --- we explain the exact meaning of this statement, and along the way we also clarify what is ``anomalous'' about the surface states of these $p=2$ semimetals. While most of the above examples have been discussed in the literature~\cite{Luttinger60,Else21,Burkov11-1,Hughes_linenode,Schnyder18,Wang20,Song21}, our work provides a unified framework to understand various different systems, and clarifies the topological aspects of the nodal line semimetals.

The rest of the paper is organized as follows. 
In Section~\ref{sec:2} we discuss odd-codimension Fermi surfaces, including ordinary metals and magnetic Weyl semimetals. In Section~\ref{sec:3} we discuss even-codimension Fermi surfaces, including graphene-like systems in 2D and nodal line semimetals in 3D. In particular, we provide a detailed discussion of the nodal line semimetals and clarify some important issues such as the anomaly of the surface states. We conclude in Section IV with a discussion of our results.

\section{Odd Fermi surface codimension: generalized chiral anomaly}
\label{sec:2}
\subsection{Ordinary metals: $p=1$}
\subsubsection{One-dimensional metals}
\label{sec:2.1}
\begin{figure}[t]
\includegraphics[width=11cm]{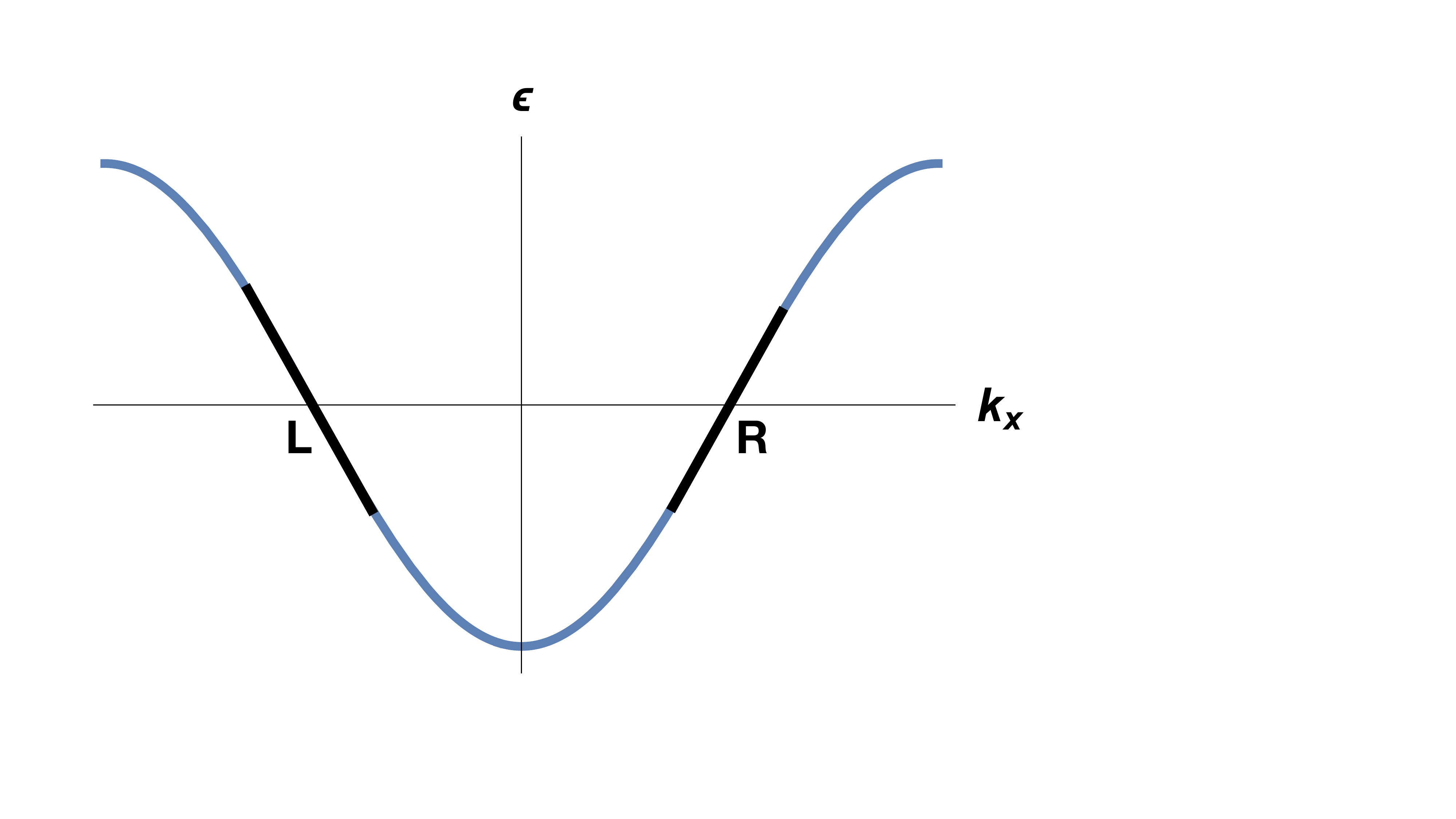}
\caption{(Color online) Band dispersion of 1D lattice fermions. Left and right-handed Weyl branches are shown in black.}
\label{fig:1}
\end{figure}

Let us start with the simplest system, a one-dimensional (1D) spinless metal. 
In this case the low-energy electronic structure, ignoring interactions for now, consists of two Fermi points at $k_{\pm} = \pm k_F$. 
Each Fermi point hosts gapless chiral modes, left (L) and right (R) handed, as depicted in Fig.~\ref{fig:1}.
The equilibrium part of the response of this 1D metal has the form 
\beq
\label{eq:1}
S = - i \int d \tau d x \,\rho A_0, 
\eeq
where $\rho$ is the equilibrium charge density, $A_0$ is the imaginary-time scalar potential
and we use $\hbar = c = e = 1$ units throughout. 
The charge density is related to the Fermi sea volume as
\beq
\label{eq:2}
\rho = \frac{2 k_F}{2 \pi} = \frac{\nu}{a}, 
\eeq
where $\nu$ is the filling (number of electrons per unit cell), introduced before, and $a$ is the lattice constant. 

To emphasize the topological character of Eq.~(1), it is useful to ``gauge'' the translation symmetry~\cite{Thorngren18,Volovik19,Nissinen21,Song21,Barkeshli21}. 
Such gauging may be viewed as a way to get rid of the dependence on a nonuniversal parameter $a$, which enters the expression for the density, focusing instead on the universal number $\nu$. 
What this means formally is that we imagine a strained crystal, in which a fixed reciprocal lattice vector $2 \pi/a$ is replaced by
\beq
\label{eq:3}
\frac{2 \pi}{a} \rightarrow \frac{2 \pi}{a}\left(\delta_{\mu x} - \partial_{\mu} u_x - x_{\mu}\right).
\eeq
Here $u_x$ is the atomic displacement and $x_{\mu}$ is an integer translation gauge field, which accounts for the fact that $u_x$ is only defined modulo the lattice constant $a$. 
The translation gauge field has two components: spatial $x_x$ and temporal $x_{\tau}$. 
Integral of $x$ over the spatial cycle gives the total number of unit cells in the 1D crystal
\beq
\label{eq:4}
\int_x x = L_x,
\eeq
where the lattice constant $a$ has been included in the integration measure $dx/a \ra dx$. 
The integral over the temporal cycle, on the other hand, defines the twist of the periodic boundary conditions in the imaginary time direction, 
i.e. the distance $\Delta L_x$, by which the crystal is shifted at $\tau = \beta = 1/T$ ($T$ is the temperature) with respect to $\tau = 0$
\beq
\label{eq:5}
\int_{\tau} x = \Delta L_x.
\eeq
Discarding the ``nontopological" pieces, Eq.~\eqref{eq:1} may then be written as~\cite{Song21,Gioia21}
\beq
\label{eq:6}
S = i \nu\, \int \epsilon_{\mu \lambda} x_{\mu} A_{\lambda} = i \nu \int \, x \wedge A.
\eeq
Note that the nonuniversal lattice constant $a$ disappears from all 
observable quantities, obtained from Eq.~\eqref{eq:6}, given the translation gauge field definitions Eqs.~\eqref{eq:4} and \eqref{eq:5}. 
Henceforth, we set the lattice constant to unity. 

Now we note that Eq.~\eqref{eq:6} may alternatively be obtained from the low-energy theory only, by invoking its emergent 't Hooft anomaly. 
Indeed, at low energies and assuming perfect translational symmetry (i.e. ignoring impurity scattering), the numbers of L and R electrons are 
separately conserved,\footnote{This remains true with electron-electron interactions as long as Umklapp scattering is irrelevant.} which corresponds to an emergent symmetry group $U(1)_L \times U(1)_R$. 
This symmetry, however, is anomalous, in the sense that it may not be realized as a true microscopic symmetry on a 1D lattice. 
In our context this simply means that the L and R quantum numbers are only defined at the corresponding Fermi points and not globally 
in the full 1D Brillouin zone (BZ), as obvious from Fig.~\ref{fig:1}. 
A 1D system with such an internal symmetry may only appear on the boundary of a 2D topological crystalline insulator, whose topological response theory 
is given by
\beq
\label{eq:CS}
S=\frac{i}{4\pi}\int(A_R \wedge dA_R-A_L\wedge dA_L).
\eeq
Each of the two terms in Eq.~\eqref{eq:CS} is a Chern-Simons term with a unit coefficient, giving rise to a pair of chiral modes of opposite chirality on the boundary --- this is just the familiar chiral anomaly in $(1+1)$ dimensions.

If we restrict the emergent $U(1)_L\times U(1)_R$ symmetry to the physical $U(1)\times \mathbb{Z}$ symmetry ($\mathbb{Z}$ being the lattice translation symmetry), we have $A_R=A+k_Fx$ and $A_L=A-k_Fx$ (recall that momentum is the ``charge'' under translation). The chiral anomaly becomes
\beqa
\label{eq:7}
S&=&\frac{i}{4 \pi} \int (A + k_F x) \wedge d (A + k_F x) \nonumber \\
&-&\frac{i}{4 \pi} \int (A - k_F x) \wedge d (A -  k_F x) = i \nu \int \, x \wedge d A, 
\eeqa
which leads to precisely Eq.~\eqref{eq:6} on the 1D boundary, assuming the translation gauge field is flat, i.e. $d x = 0$.
The reason Eq.~\eqref{eq:6} actually exists in a stand-alone 1D system is that the true symmetry of the 1D metal is not $U(1) \times U(1)$, 
but $U(1) \times \mathbb{Z}$, where $\mathbb{Z}$ corresponds to discrete translations. 
This is reflected in the fact that $x$ in Eq.~\eqref{eq:6} is not a $U(1)$ gauge field, but a discrete translation gauge field. 
Moreover, the translational symmetry is not an internal symmetry. 

These considerations reveal the connection between the Luttinger theorem in a 1D metal and the 't Hooft anomaly of the emergent 
$U(1)_L \times U(1)_R$ symmetry of its low-energy theory. 
In what follows, we generalize these arguments to ordinary metals in higher dimensions, as well as to topological semimetals. 
\subsubsection{Two-dimensional metals}
\label{sec:2.2}
Let us now generalize the above ideas to higher-dimensional metals. For simplicity we consider the 2D case only, 
since it contains all of the essential new features that distinguish higher-dimensional metals from the 1D case. 
In 2D the Fermi surface is a continuous 1D manifold. Generalizing the emergent chiral charge conservation to this case, one assumes 
independent charge conservation at every point on the Fermi surface, which is in fact the standard assumption of the Fermi liquid theory.
This implies that the emergent symmetry group is $LU(1)$, the so-called loop group of $U(1)$~\cite{Else21}.
Assuming a single Fermi surface sheet for simplicity, the 1D Fermi surface acts, in effect, as an extra dimension. 
Let us denote the coordinate, which parametrizes the Fermi surface as $\theta \in [0, 2 \pi]$, and let $k_{x,y}(\theta)$ be the corresponding 
components of the Fermi momentum. 
The total ``space-time" dimension of this system, which includes the extra $\theta$-dimension, is $2 + 1 + 1 = 4$. 
The 2D Fermi surface may thus be connected to the 't Hooft anomaly of a $(4+1)$-dimensional topological insulator, 
described by the following  Chern-Simons topological field theory~\cite{Else21, MaWang2021}
\beq
S=\frac{i}{6(2\pi)^2}\int \A \wedge d\A \wedge d\A,
\eeq
where $\A$ is the probe gauge field living in both real and (Fermi surface) momentum space. Accordingly, the exterior derivative operation $d$ above contains derivatives with respect to the spacetime coordinates (including the extra space dimension), as well as the Fermi surface coordinate $\theta$. 

Following previous logic, we now restrict to the physical $U(1)\times\mathbb{Z}^2$ symmetry (charge conservation and two lattice translations), and set $\A=A+k_xx+k_yy$. The $LU(1)$ anomaly becomes
\beqa
\label{eq:8}
S&=&\frac{i}{6 (2 \pi)^2} \int (A + k_x x + k_y y) \nonumber \\
&\wedge&d (A + k_x x + k_y y) \wedge d (A + k_x x + k_y y), 
\eeqa
where $x, y$ are gauge fields, corresponding to the translational symmetry in the $x$ and $y$-directions respectively. 
The nonvanishing part of this, which gives rise to the theory of a 2D Fermi liquid, is the following mixed anomaly term
\beq
\label{eq:9}
S = \frac{i}{2 (2 \pi)^2} \int (k_x x + k_y y) \wedge d (k_x x + k_y y) \wedge d A. 
\eeq
Assuming $d x = d y = 0$, as before, this corresponds to the following boundary theory
\beq
\label{eq:10}
S = \frac{i}{2 (2 \pi)^2} \int (k_x \partial_{\theta} k_y - k_y \partial_{\theta} k_x) x \wedge y \wedge A.
\eeq
Taking into account that 
\beq
\label{eq:11}
\frac{1}{2} \int_0^{2 \pi} d \theta (k_x \partial_{\theta} k_y - k_y \partial_{\theta} k_x) = V_F, 
\eeq
where $V_F$ is the 2D Fermi sea volume, we obtain the following $2 + 1$-dimensional topological field theory
\beq
\label{eq:12}
S = \frac{i V_F}{(2 \pi)^2} \int x \wedge y \wedge A = i \nu \int x \wedge y \wedge A, 
\eeq
which, similar to Eq.~\eqref{eq:6}, declares that the charge per unit cell is given by $\nu=V_F/(2\pi)^2$. This is precisely the field theory expression of the Luttinger theorem in a 2D Fermi liquid~\cite{Song21}.

\subsection{Magnetic Weyl semimetal: $p=3$}
\label{sec:3.1}

Now we demonstrate that identical arguments may be used to describe 3D topological semimetals. Topological semimetals exist at integer fillings, which would normally correspond to insulators.
This means that the filling anomaly, which corresponds to the Luttinger theorem in Fermi liquids, does not exist in this case. 
However, just as ordinary metals may be viewed as intermediate phases between insulators, corresponding 
to different integer values of the filling $\nu$, topological semimetals may be viewed as intermediate gapless phases between 
pairs of topologically-distinct insulators at the {\em same} fixed integer filling. 
Correspondingly, in each case there exists a continuously-variable parameter, which tunes between the two insulators and 
acts as the coefficient of the corresponding anomaly term, analogously to the filling $\nu$ in Fermi liquids. 
While many different kinds of topological semimetals exist, here we  focus on the magnetic Weyl semimetal. In Sec.~\ref{sec:3.2} we discuss nodal line semimetals in detail. For an in-depth discussion of other types of topological semimetals from a similar viewpoint, see Ref.~\cite{Gioia21}. 

A magnetic Weyl semimetal has an odd number of pairs of opposite-chirality Weyl nodes at the Fermi energy. 
We consider the simplest case when this number is one and assume the nodes are separated along the 
$z$-axis. 
Let the node coordinates be $k_z^{\pm} = \pm Q \hat z$. 
The separation between the nodes $2 Q$ is a continuously-tunable parameter.
$2 Q = 0$ corresponds to an ordinary magnetic insulator with zero Hall conductivity, 
while $2 Q = 2 \pi$ (i.e. a reciprocal lattice vector) corresponds to a quantum anomalous Hall insulator with a
Hall conductivity
\beq
\label{eq:13}
\sigma_{xy} = \frac{1}{2 \pi} \frac{2 Q}{2 \pi} = \frac{1}{2 \pi}, 
\eeq
which is equivalent to a Hall conductance quantum $1/2 \pi$ per $xy$ atomic plane, i.e. this is a weak topological insulator that may be viewed 
as a stack of 2D integer quantum Hall insulators. 
Intermediate values of $2 Q \in (0, 2 \pi)$ correspond to the Weyl semimetal. 
In this case the Hall conductance per atomic plane $\sigma_{xy} = 2 Q /4 \pi^2$ takes a noninteger value, which is impossible in an insulator, 
unless it has topological order. This is the analog of a fractional filling in a Fermi liquid metal. 

The corresponding emergent anomaly is the familiar chiral anomaly of the emergent $U(1)_L\times U(1)_R$ symmetry corresponding the the charge conservation at each Weyl node. If the $U(1)_L\times U(1)_R$ symmetry is exact, the system can only be realized on the boundary of a 4D topological insulator, described by the following $(4 + 1)$-dimensional Chern-Simons theory
\beq
\label{eq:WeylAnomaly}
S=\frac{i}{6(2\pi)^2}\int (A_R\wedge dA_R \wedge dA_R - A_L\wedge dA_L\wedge dA_L).
\eeq

We now restrict to the physical $U(1)$ and $z$-translation symmetry, and make the substitution $A_R=A+Qz$, $A_L=A-Qz$. The anomaly becomes
\beqa
\label{eq:14}
S&=&\frac{i}{6 (2 \pi)^2} \int (A + Q z) \wedge d (A + Q z) \wedge d(A + Q z) \nonumber \\
&-&\frac{i}{6 (2 \pi)^2} \int (A - Q z) \wedge d (A - Q z) \wedge d(A - Q z) \nonumber \\
&=&\frac{i Q}{4 \pi^2} \int z \wedge d A \wedge dA. 
\eeqa
This gives a $(3+1)$-dimensional boundary theory
\beq
\label{eq:15}
S = \frac{i Q}{4 \pi^2} \int z \wedge A \wedge d A,
\eeq
which describes the topological response of the magnetic Weyl semimetal. 
Indeed, Eq.~\eqref{eq:15} encodes precisely the Hall conductance of 
\beq
\label{eq:16}
G_{xy} = \frac{2 Q}{4 \pi^2} L_z.
\eeq

There exists a close connection between the magnetic Weyl semimetal and the 1D metal, discussed in Section~\ref{sec:3.1}. 
This may be seen by inserting a magnetic flux line carrying a unit flux quantum $\Phi = 2 \pi$ in the $xy$-plane. 
According to Eq.~\eqref{eq:15}, this flux line is described by the action
\beq
\label{eq:17}
S = \frac{i 2 Q}{2 \pi} \int z \wedge A. 
\eeq
Comparing this with Eq.~\eqref{eq:6}, we see that the $2 \pi$ flux line carries a 1D metal, with the effective filling $\nu = 2 Q/ 2 \pi$. 
This is just another way to arrive at the familiar result that a Weyl semimetal in an external magnetic field has the lowest Landau level that 
crosses the Fermi energy at the location of the Weyl nodes.

We note that we can derive the emergent anomaly of a magnetic Weyl semimetal from that of a three-dimensional metal. The latter is described by a $7$-dimensional [$(3+1)$ from real spacetime, $2$ from Fermi surface momentum, and $1$ extra ``bulk" dimension] Chern-Simons term
\beq
\label{eq:7dcs}
S=\frac{i}{6(2\pi)^3}\int \A \wedge d\A \wedge d\A \wedge d\A.
\eeq
A Weyl fermion can be recovered in the limit of a small spherical Fermi surface, with a total Berry curvature $\int_{FS}d\mathcal{A}=2\pi$ on the Fermi surface. This reduces Eq.~\eqref{eq:7dcs} to a $5$-dimensional Chern-Simons term, which is exactly the anomaly of a Weyl fermion Eq.~\eqref{eq:WeylAnomaly}. This procedure can be straightforwardly generalized to higher dimensions, from which we can obtain the emergent anomaly of a general odd-codimension Fermi surface, as we discuss next.

\subsection{General dimensions}

Using the (momentum space) dimensional reduction scheme described above, the pattern becomes apparent for general space dimension $d$ and Fermi surface dimension $d-p$, with $p$ odd. The emergent symmetry corresponds to charge conservation at each point on the $d-p$-dimensional Fermi surface. The probe gauge field $\A$ then lives in both real space-time, as well as the $d-p$-dimensional momentum space. The corresponding 't Hooft anomaly can be described by a Chern-Simons term in $(d+1)+(d-p)+1=2d-p+2$ dimension (note the importance of $p$ being odd):
\beq
S=\frac{i}{N_{d,p}}\int \A(d\A)^{d-(p-1)/2},
\eeq
where all the products in the integrand are wedge products, and the normalization factor
\beq
N_{d,p}=(d-(p-3)/2)!(2\pi)^{d-(p-1)/2}.
\eeq 

If we restrict to the physical $U(1)$ and translation symmetries (with gauge fields represented by $A$ and $x_1,x_2...$), the anomaly term becomes a response term in the physical $d+1$ space-time dimensions
\beq
S=\frac{iV^F_{i_1...i_{d+1-p}}}{(\frac{p+1}{2})!(2\pi)^{d-(p+1)/2}}\int A(dA)^{(p-1)/2}x_{i_1}x_{i_2}...x_{i_{d+1-p}},
\eeq
where $V^F_{i_1...i_{d+1-p}}$ is the Fermi surface volume, projected to the $i_1...i_{d+1-p}$ subspace. Physically this represents a nontrivial Chern-Simons type of response per unit length scale.

\section{Even Fermi surface codimension: generalized parity anomaly}
\label{sec:3}

\subsection{Nodal line semimetal: $p=2$}
\label{sec:3.2}
\subsubsection{Topological response}
We now discuss the case of nodal line semimetals. 
Since this has not received as much attention in the literature as point node semimetals, we discuss this case in more detail. 
Nodal line semimetals with nontrivial topology arise only when bands are nondegenerate~\cite{Burkov11-2,Hughes_linenode}.
This requires breaking of either time reversal or inversion symmetry and nonvanishing spin-orbit coupling. 
The gaplessness of the nodal line is then protected by mirror reflection symmetry in the plane, containing the nodal line. 
Given this, we adopt a simple two-band model, with the following cubic-lattice Hamiltonian~\cite{Wang17,Shapourian18}
\beqa
\label{eq:18}
{\cal H}(\bk)&=&\left[6 - t_1 - 2 (\cos k_x + \cos k_y + \cos k_z) \right] \sigma_x \nonumber \\
&+&2 t_2 \sin k_z \sigma_y + t_3 (2 - \cos k_x - \cos k_y). 
\eeqa
The nodal line in this model appears in the $xy$-plane and is protected by the mirror reflection symmetry within this plane, where 
the mirror reflection operator is $\sigma_x$. 
It is implicit that the time reversal symmetry is broken, since the bands are nondegenerate. 
Exactly this sort of Hamiltonian describes the low-energy electronic structure in, for example, the magnetic multilayer model of Ref.~\cite{Burkov11-2}.

In the absence of the last term in Eq.~\eqref{eq:18}, proportional to the unit matrix, the Hamiltonian has an exact particle-hole symmetry, which 
requires all points on the nodal line to be at the same energy. However, once the last term is included, this is no longer the case and the electronic 
structure consists of electron and hole pockets of zero total Luttinger volume. We first derive the topological response of the nodal line assuming 
the particle-hole symmetry and then argue that relaxing this assumption does not change the response. 

Following our logic in the previous examples, we consider an emergent symmetry $LU(1)$, that corresponds to charge conservation at each point on the nodal line. This is justified by the fact that all weak short-range interactions, including attractive, are irrelevant in this case due to the vanishing density of states. 
The corresponding probe gauge field $\A$ lives in $(3+1)$-dimensional real space-time as well as $1$-dimensional momentum space. If there is an emergent anomaly, associated with this symmetry, it is a topological term in one higher dimension. The most obvious such term in $(3+1)+1+1=6$ dimensions is the $\theta$-term with $\theta=\pi$:
\beq
\label{eq:6dparityanomaly}
S=\pm\frac{i\pi}{6(2\pi)^3}\int d\A \wedge d\A \wedge d\A,
\eeq
where the arbitrariness of the sign reflects the $2 \pi$ periodicity of the coefficient. The quantization of the $\theta$ angle comes from the mirror reflection symmetry, which inverts only one coordinate (the real space $z$) and sends $\theta\to -\theta$. This is consistent with the fact that the nodal ring can be trivially gapped if the reflection symmetry is broken.

The above generalized parity anomaly can in fact be derived from the generalized chiral anomaly of ordinary metals, using a dimensional-reduction scheme similar to that in Sec.~\ref{sec:3.1}. To see this, consider a slightly electron-doped nodal line, such that the Fermi energy $\epsilon_F > 0$ (still assuming particle-hole symmetry as in the undoped state). In this case we have a toric Fermi surface, which encloses the nodal line along its length. As discussed originally in Ref.~\cite{Else21} and reviewed in Sec.~\ref{sec:2.2}, the Fermi surface carries an emergent anomaly, encoded by the $(6+1)$-dimensional Chern-Simons term
\beq
\label{eq:7dchiralanomaly}
S=\frac{i}{6(2\pi)^3}\int \A \wedge d\A \wedge d\A \wedge d\A.
\eeq

A characteristic feature of the nodal line is the $\pi$ Berry phase, accumulated along a path in momentum space, enclosing the nodal line~\cite{Burkov11-2}. Let the toric Fermi surface be parametrized by two angular variables $\theta, \phi \in [0, 2\pi]$, where $\theta$ refers to the azimuthal direction, 
which traverses the Fermi surface without enclosing the nodal line, while $\phi$ is the polar direction, which wraps around the nodal line. Then we have
\beq
\label{eq:24}
\int_0^{2 \pi}  d \phi \A_{\phi} = \pm \pi. 
\eeq
Plugging this into Eq.~\eqref{eq:7dchiralanomaly} and taking the limit $\epsilon_F \rightarrow 0$, we again arrive at Eq.~\eqref{eq:6dparityanomaly}.

Following 
the same logic as in the previous cases, we can now restrict to the microscopic symmetries [$U(1)$, lattice translations and the mirror reflection $z\to-z$]. Introducing a parameter $\theta \in [0, 2 \pi]$, such that $k_{x,y}(\theta)$ are the coordinates of the points on the nodal line, we have $\A=A+k_xx+k_yy$.   
The $6$-dimensional $\theta$-term now becomes
\beqa
\label{eq:19}
S&=&\pm \frac{i \pi}{6 (2 \pi)^3} \int d (A + k_x x + k_y y) \wedge d (A + k_x x + k_y y) \nonumber \\
&\wedge& d (A + k_x x + k_y y).
\eeqa
 This gives rise to a mixed anomaly term, which is similar to Eq.~\eqref{eq:9}, except in one extra dimension
\beq
\label{eq:20}
S = \pm \frac{i \pi}{2 (2 \pi)^3} \int d (k_x x + k_y y) \wedge d (k_x x + k_y y) \wedge d A,
\eeq
corresponding to a boundary theory
\beqa
\label{eq:21}
S&=&\pm \frac{i \pi}{2 (2 \pi)^3} \int (k_x \partial_{\theta} k_y - k_y \partial_{\theta} k_x) x \wedge y \wedge d A \nonumber \\
&=&\pm \frac{i V_F}{8 \pi^2} \int x \wedge y \wedge d A, 
\eeqa
where $V_F$ is the area in momentum space, enclosed by the nodal line. 

The physical meaning of Eq.~\eqref{eq:21} is that it describes spontaneous (i.e. existing in the absence of an external electric field) electric polarization~\cite{Hughes_linenode,Song21}, given by
\beq
\label{eq:22}
P = \pm \frac{V_F}{8 \pi^2}. 
\eeq
The two signs in front reflect the fact that, in the presence of the mirror symmetry, there must exist two degenerate values of polarization,
related to each other by the mirror reflection (unless $P = 0$). 
If the nodal line is expanded to the edges of the BZ, in which case a gap opens, we have $V_F = (2 \pi)^2$ and $P = \pm 1/2$. 
This half-quantized polarization is the only nontrivial value of polarization that an inversion (or mirror) symmetric insulator can have.
A nontrivial ``fractional" value of polarization in Eq.~\eqref{eq:22} means that the polarization is not uniquely defined, as manifested by the arbitrary sign in Eq.~\eqref{eq:22}, which in turn requires a gapless spectrum, i.e. the nodal line. 

The precise meaning of polarization Eq.~\eqref{eq:22} for a gapless system requires a more careful explanation. Unlike the Chern-Simons type of responses (such as the Hall conductivity or the charge density), which can be easily defined even for gapless systems, the $\theta$-term type of responses are sometimes ill-defined in this case. One well-known example is the 1D metal, which does not have a well-defined polarization (this is a direct consequence of the $(1+1)$d chiral anomaly). Similarly, the magnetoelectric polarizability (the $dAdA$ $\theta$-angle) of a 3D Dirac semimetal is not defined. So what exactly do we mean by Eq.~\eqref{eq:22} for the nodal line semimetal? One possible answer is that if we weakly break the mirror reflection symmetry and completely gap out the nodal line, the resulting insulator has a spontaneous polarization given by Eq.~\eqref{eq:22}, with the overall sign determined by the sign of the mirror symmetry breaking. However, one may still ask if we can associate the polarization directly to some observable in the semimetal phase, instead of having to rely on neighboring phases to define it (such as mirror-breaking insulators). 

The definition of polarization through the Luttinger theorem violation on the boundary~\cite{Song21} turns out to be useful here. If the surface state is a 2D Fermi liquid at low energy, the bulk polarization is given by
\beq
\label{eq:PolarizationLuttinger}
P=\rho-\frac{V_F}{(2\pi)^2} \hspace{5pt} {\rm{mod}}\hspace{2pt}1,
\eeq
where $P$ is the polarization density, perpendicular to the surface, $\rho$ is the surface charge density (per unit cell) and $V_F$ is the Luttinger volume of the surface Fermi liquid. One can think of the first term as the classical contribution (which interprets polarization as a dipole moment) and the second term as the quantum correction.~\footnote{If the surface is not a Fermi liquid, we simply replace the second term by the appropriate ``anomaly indicator"~\cite{Song21}, which is essentially the charge density required to realize the state in a purely two-dimensional system.}
\begin{figure}[t]
\includegraphics[width=\linewidth]{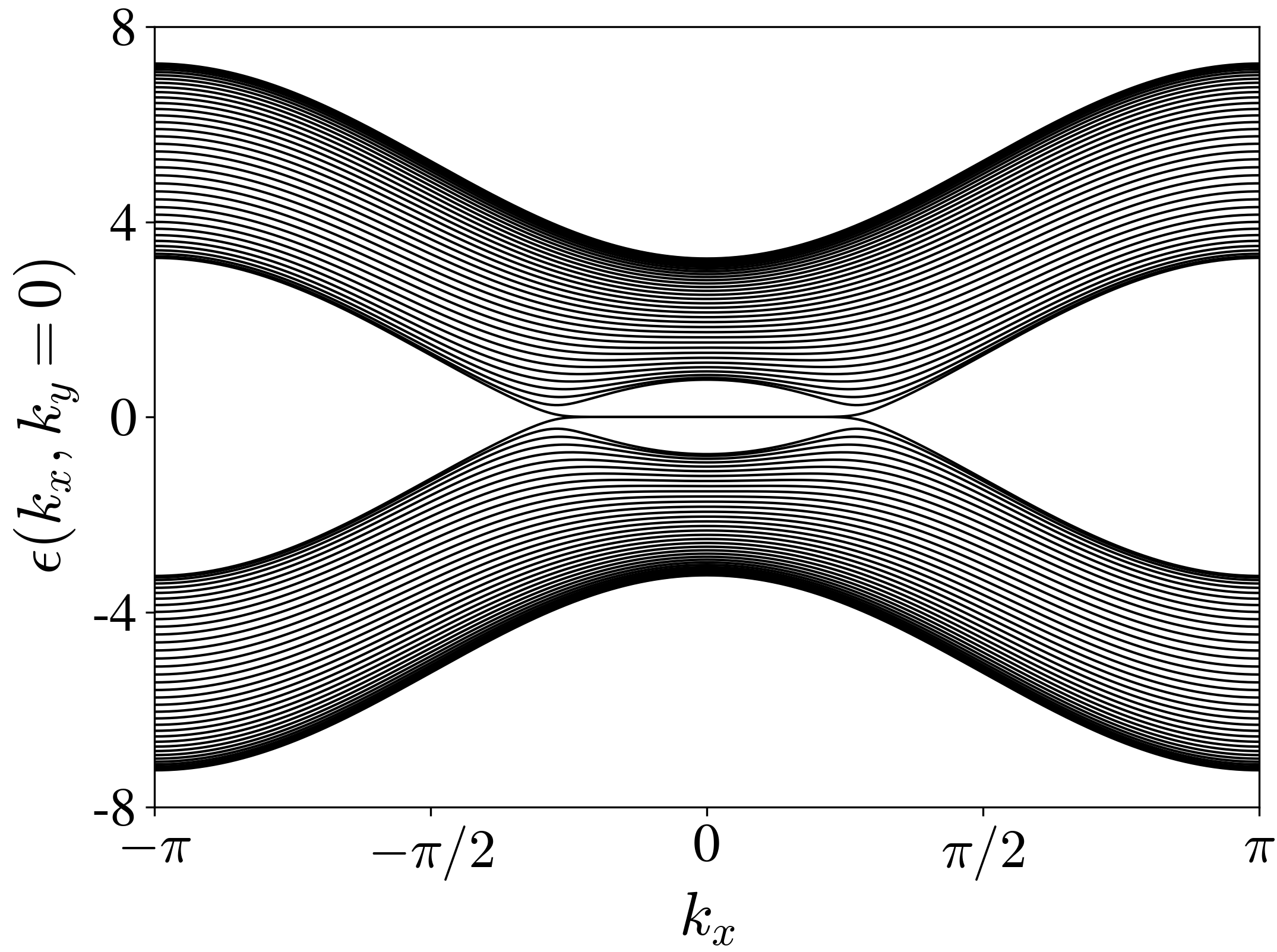}
\caption{Electronic structure of a finite size nodal line system, described by Eq.~\eqref{eq:18} with $t_1 = 0.75, t_2 = 1, t_3 = 0$. The perfectly flat surface state acquires a dispersion if $t_3 \neq 0$.}
\label{fig:drumhead}
\end{figure}

The band structure for an $xy$-surface of the nodal line semimetal is schematically illustrated in Fig.~\ref{fig:drumhead}. The most important feature is a branch of fermion modes that exist only inside the region, enclosed by the projection of the nodal line onto the surface BZ. Other features, such as the flatness of these modes, are less robust and do not play a role in the following discussion. Now let us put the surface chemical potential somewhere and calculate the polarization from Eq.~\eqref{eq:PolarizationLuttinger}. The answer, however, depends on the surface chemical potential. We denote the polarization, calculated when the chemical potential is above (below) zero, which is taken as the energy of the nodal line, as $P_+$ ($P_-$). We find
\beq
P_+-P_-=\frac{V_F}{4\pi^2}.
\eeq
This means that the bulk state cannot have a uniquely defined polarization (unless $V_F=4\pi^2$, which is trivial), and therefore cannot be a short-range entangled insulator. Since the reflection symmetry inverts polarization, we must have
\beq
P_+=-P_-=\frac{V_F}{8\pi^2},
\eeq
which is exactly Eq.~\eqref{eq:22}.

The above discussion also clarifies the anomalous nature of the surface states of the nodal line semimetal: while a surface Fermi liquid per se appears to be quite non-anomalous (just like any other 2D metal), the Luttinger theorem is violated. Such Luttinger theorem violation also happens on the surface of a ferroelectric insulator [Eq.~\eqref{eq:PolarizationLuttinger}], but the violation on the surface of a nodal line semimetal is stronger in the sense that it cannot be removed by redefining the surface charge density by a constant shift $P$.

We now consider breaking the particle-hole symmetry, so that the energy is no longer constant on the nodal line. We ask how the above discussion of polarization could change. Again, consider moving the surface chemical potential gradually from well below the nodal line to well above. The difference now is that there are intermediate regions, where the chemical potential is above only parts of the nodal line. The resulting Fermi liquid has open Fermi surfaces (arcs) and $V_F$ cannot be defined. As a result, the polarization $P$ cannot be defined in the intermediate region. However, when $P$ is well defined (when the chemical potential is far away from the nodal ring), it still takes the values $P_{\pm}=\pm V_F/8\pi^2$.

\begin{figure}[t]
\includegraphics[width=\linewidth]{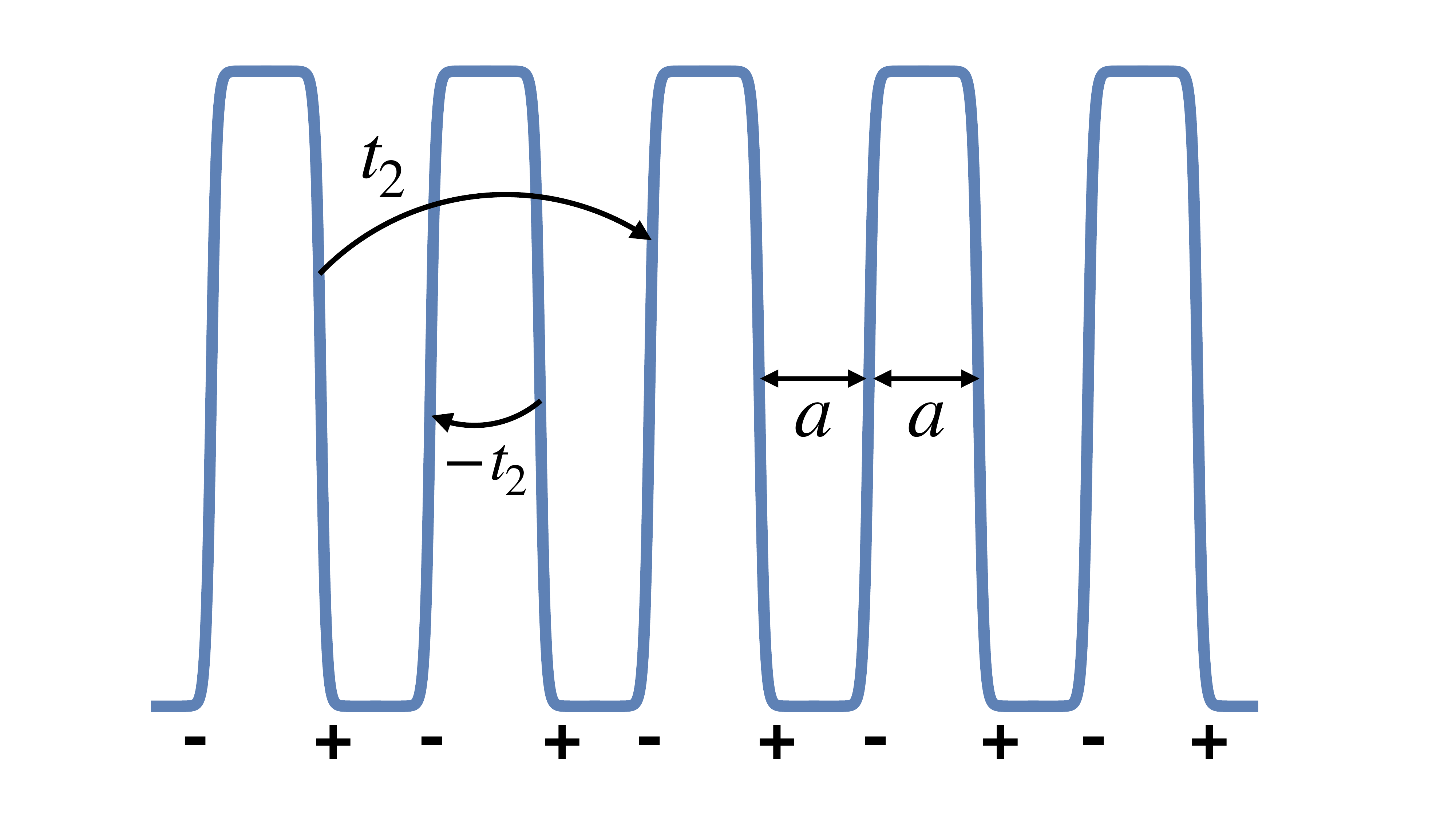}
\caption{(Color online) Stacked domain wall construction. Each domain wall, separating regions with different sign of the mirror symmetry breaking ``mass" $m$, hosts a spinless particle or hole-like Fermi liquid state. The $\pm$ signs are the corresponding eigenvalues of $\sigma_x$.}
\label{fig:2}
\end{figure}

\subsubsection{Alternative perspective: Coupled-layer derivation of the topological response}

We can also derive the nodal line topological response Eq.~\eqref{eq:21} by an entirely different, more microscopic method. 
This method allows us to demonstrate more explicitly that particle-hole symmetry is not essential and the response is unchanged when it is broken. 
We start with the nodal line Hamiltonian Eq.~\eqref{eq:18} (but still assuming particle-hole symmetry for now) and add a mirror symmetry breaking 
``mass term"
\beqa
\label{eq:25}
&&{\cal H}(\bk) = \left[6 - t_1 - 2 (\cos k_x + \cos k_y + \cos k_z) \right] \sigma_x \nonumber \\
&+&2 t_2 \sin k_z \sigma_y + m \sigma_z.
\eeqa
Let us now consider a domain wall, at which the mass $m$ changes sign from negative to positive as a function of $z$. 
As in other, more familiar cases, topologically nontrivial nature of the gapless nodal line state manifests in a gapless bound state 
on such a domain wall.
To find this state, we replace $k_z\rightarrow-i\partial/\partial_z$, keeping only up to linear order terms in $k_z$. Inclusion of the higher order terms does not affect the gaplessness of the bound state~\cite{PhysRevB.101.081102}. We thus consider the following real-space Hamiltonian
\beqa
\label{eq:26}
&&{\cal H}(k_x,k_y,z) = (4 - t_1 - 2 \cos k_x - 2 \cos k_y) \sigma_x \nonumber \\
&-&2 i t_2 \sigma_y \frac{\partial}{\partial z} + m(z) \sigma_z,
\eeqa
where $m(z \ra - \infty) < 0,\, m(z \ra \infty) > 0$. 
A straightforward and standard calculation gives the wavefunction of the bound state as
\beq
\label{eq:27}
\Psi_-(z) = e^{- \frac{1}{2 t_2} \int_0^z d z' m(z')} |\sigma_x = - 1\rangle, 
\eeq
and the corresponding energy eigenvalue 
\beq
\label{eq:28}
\epsilon_-(k_x, k_y) = - (4 - t_1 - 2 \cos k_x - 2 \cos k_y). 
\eeq
A domain wall of opposite topological charge, where $m$ changes sign from positive to negative, binds a state with 
\beq
\label{eq:29}
\Psi_+(z) = e^{- \frac{1}{2 t_2} \int_0^z d z' m(z')} |\sigma_x = 1\rangle, 
\eeq
and 
\beq
\label{eq:30}
\epsilon_+(k_x, k_y) = 4 - t_1 - 2 \cos k_x - 2 \cos k_y. 
\eeq
These bound states correspond to spinless 2D electron or hole-like Fermi liquids. 
Their topological responses are thus given by Eq.~\eqref{eq:12}, where the coefficient is the corresponding 
Luttinger volume, including opposite signs in the electron and hole cases. 
Note that these domain wall states by themselves are not anomalous, in the sense that a purely 2D lattice realization clearly exists. 

Now we note that the gapless nodal line state may be reconstructed by stacking alternating domain walls with opposite topological charges, while preserving mirror symmetry, as shown in Fig.~\ref{fig:2}. 
To preserve the mirror symmetry, the inter-domain wall tunneling Hamiltonian must have the form
\beq
\label{eq:31}
{\cal H}_t(k_z) = 2 t_2 \sigma_y \sin k_z, 
\eeq
i.e. exactly the form of the second term in Eq.~\eqref{eq:25}, which corresponds to inter-unit cell tunneling with a direction-dependent 
sign of the tunneling matrix element.  

To obtain the topological response of the nodal line we need to sum the contributions from the domain wall bound states
\beq
\label{eq:32}
S = \pm i \frac{V_F}{(2 \pi)^2}\int \sum_n (-1)^n A_0(z_n),
\eeq
where $V_F$ is the magnitude of the Luttinger volume of each 2D domain-wall-bound Fermi liquid, $z_n$ is the coordinate 
of the $n^\text{th}$ domain wall and the arbitrariness of the overall sign reflects the arbitrariness of assigning the starting point of the summation. 
Taking the continuum limit and switching to the translation gauge field notation, following the logic of Section~\ref{sec:2.1}, we arrive at 
the topological response theory of Eq.~\eqref{eq:21}.

Now let us add the particle-hole symmetry breaking term $t_3 (2 - \cos k_x - \cos k_y) - \epsilon_F$ to the nodal line Hamiltonian.
A finite Fermi energy $\epsilon_F$ is necessary to keep charge neutrality once the particle-hole symmetry is broken. 
It is clear that the domain wall bound states remain the same, except for a shift in their energies
\beqa
\label{eq:33}
\epsilon_{\pm}(k_x, k_y)&=&t_3 (2 - \cos k_x - \cos k_y) - \epsilon_F \nonumber \\
&\pm&(4 - t_1 - 2 \cos k_x - 2 \cos k_y),
\eeqa
i.e. the Luttinger volumes of the electron and hole-like domain wall states get changed by equal and opposite amounts, so that 
overall charge neutrality is preserved. 
It follows that the topological response term Eq.~\eqref{eq:32} remains unchanged. 

\subsubsection{Gapped topological orders}
The above picture of the nodal line as a stack of 2D Fermi liquid domain wall states offers a useful prospective on the question of gapping the 
nodal line while preserving its anomalous response Eq.~\eqref{eq:21}. 
Following the ``vortex condensation" approach, originally applied to surface states of 3D topological insulators~\cite{Wang13,Metlitski15} and 
later generalized to 3D topological semimetals in Refs.~\cite{Wang20,Thakurathi20,Sehayek20} (also see Ref.~\cite{Kane21} for related work), we may start with a gapped superconducting 
nodal line state and then attempt to construct an insulator with the same topological response by condensing vortices in the superconductor. 
A simple mean-field picture of this state may be obtained by applying this procedure to each 2D domain wall state independently. 
Since each domain wall hosts a spinless 2D Fermi liquid, the simplest fully gapped superconducting state is $p$-wave, described by the following 
Hamiltonian
\beq
\label{eq:34}
H = \sum_{\bk} \left[\epsilon_{\pm}(\bk) c^\dg_{\bk}c^\pdg_{\bk} + \frac{\Delta}{2} (\sin k_x + i \sin k_y) c^\dg_{\bk} c^\dg_{-\bk} + \text{h.c.}\right],
\eeq
where $\epsilon_{\pm}(\bk)$ are given by Eqs.~\eqref{eq:28}, \eqref{eq:30} (we ignore the particle-hole symmetry breaking here for simplicity) 
and $\bk$ is the 2D crystal momentum. 
Introducing Nambu spinor notation $\psi_{\bk} = (c^\pdg_\bk, c^\dg_{- \bk})$, this may be represented as a massive 2D Dirac Hamiltonian
\beq
\label{eq:35}
H = \frac{1}{2} \sum_\bk \psi^\dg_\bk \left[\epsilon_{\pm}(\bk) \tau_z + \Delta (\tau_x \sin k_x - \tau_y \sin k_y) \right] \psi^\pdg_\bk, 
\eeq
where $\tau_a$ are Pauli matrices in the particle-hole space. 
This representation makes it clear that the superconducting domain wall state is a Read-Green topological superconductor~\cite{Read00},
which hosts chiral Majorana modes at the edges, with chirality determined by the type (electron or hole-like) of the domain wall state. 
Consequently, an elementary flux $h c/2 e = \pi$ vortex hosts a zero-energy localized Majorana bound state and can not be condensed. 
A double, or flux $2 \pi$, vortex does not host any zero-energy states, but may still not be condensed without breaking symmetries. 
This is a consequence of the fact that such vortices experience the nontrivial band filling $\pm \nu$ of the 2D domain wall Fermi liquid (the signs 
distinguish between the electron and hole-like states) as 
magnetic flux, leading to a projective representation of the group of translations every time a vortex line intersects with a layer~\cite{Balents05}.
\beq
\label{eq:36}
T_x T_y = e^{\pm 2 \pi i \nu} T_y T_x. 
\eeq
This effect is most obvious when the system has a boundary, in which case the sign in Eq.~\eqref{eq:36} depends on exactly how the system is terminated. This, in turn, means that $2 \pi$ flux vortices carry nontrivial momentum and their condensation leads to breaking of the crystal translational 
symmetry. 
When the filling $\nu$ is rational, i.e. $\nu = p/q$ with $p$ and $q$ mutually prime integers, flux $2 \pi q$ vortices may be condensed without breaking 
any symmetries. This leads to an insulator with $\mathbb{Z}_q$ topological order, 
which reproduces the response of the gapless nodal line state. This $\mathbb{Z}_q$ topological order contains both particle-like and loop-like excitations. The loop excitations are the remnants of the uncondensed vortices, with the property that each time the loop intersects with a layer, the translation symmetry action changes by Eq.~\eqref{eq:36}. Similar ``foliated" structures for the loop excitations have been discussed  by some of us in the closely related context of a gapped magnetic Weyl semimetal in 
Refs.~\cite{Wang20,Thakurathi20,Sehayek20}. We defer a detailed analysis of the foliated 3D topological field theory for gapped nodal line semimetals to future work. 

\subsection{General dimensions}

We can now easily generalize to arbitrary space dimension $d$ with an even codimension $p$ for the Fermi surface. The generalized parity anomaly lives in $(d+1)+(d-p)+1=2d-p+2$ dimensions and is described by a $\theta$-term at $\theta=\pi$:
\beq
\label{eq:generalparity}
S=\frac{i\pi}{N_{d,p}}\int (d\A)^{d+1-p/2},
\eeq
where all the products in the integrand are wedge product, and the normalization factor
\beq
N_{d,p}=(d+1-p/2)!(2\pi)^{d+1-p/2}.
\eeq

Unlike the cases with odd $p$, the parity anomaly requires extra discrete symmetries, such as reflection, to quantize the $\theta$-angle at $\pi$.

When restricted to the microscopic $U(1)$ and translation symmetries, Eq.~\eqref{eq:generalparity} becomes
\beq
S=\frac{iV^F_{i_1...i_{d+1-p}}}{(\frac{p}{2})!(2\pi)^{d-p/2}}\int (dA)^{p/2}x_{i_1}x_{i_2}...x_{i_{d+1-p}},
\eeq
where $V^F_{i_1...i_{d+1-p}}$ is the Fermi surface volume projected to the $i_1...i_{d+1-p}$ subspace. Physically this represents a nontrivial $\theta$-term type of response per unit length scale.

Another familiar case is when $d=p=2$, which is nothing but graphene-like Dirac semimetals in 2D, with the emergent anomaly being the standard parity anomaly for each Dirac cone. For example, consider a spinless system with two Dirac cones, separated by momentum $Q\hat{x}$, protected by reflection symmetry $y\to-y$. It is straightforward to repeat the analysis for the nodal line semimetal and conclude that the polarization in the $\hat{y}$ direction is $P_y=\pm Q/2$. In fact the situation is even simpler than the nodal line case, since polarization can be defined directly in the bulk without having to go to the boundary, as discussed in Ref.~\cite{Song21}.

\section{Discussion and conclusions}
\label{sec:4}
The main goal of this paper was to introduce a common framework, which may be used to describe topological properties of both ordinary 
metals and topological semimetals. 
We have demonstrated that both may be described in terms of unquantized (i.e. having continuously-variable coefficients) anomalies, 
which in turn may be connected to quantized topological response terms of higher-dimensional SPT insulators, associated with the emergent symmetries of the (semi)metal. 
The unquantized coefficients arise from tunable charges of crystal symmetry gauge fields (e.g. crystal momentum in the case of translational symmetry). 

Such a common framework is useful, in part, because it emphasizes similarities between ordinary and topological metals. In both cases
the necessary existence of gapless modes may be connected to topology, a viewpoint advocated early on by Volovik~\cite{Volovik03,Volovik07}.
It also compactly encodes the physical properties of metals, which are directly connected to topology. 
This may be viewed as a way to generalize Luttinger's theorem to (almost) all bulk metals. 

One concrete result along these lines is an improved understanding of nodal line semimetals in three dimensions. In particular, we have shown that the area, enclosed by the nodal line, determines the electric polarization of the semimetal. By carefully defining the concept of polarization for such semimetals, we have also clarified the anomalous nature of the ``drumhead" surface states of nodal line semimetals.

One specific type of a metal this scheme omits is the so-called type-II Dirac semimetal~\cite{Kane12,Steinberg14,Nagaosa14,Parameswaran13}.
The electronic structure of these materials contains Dirac points at time reversal invariant momenta (TRIM) at the edge of the BZ, terminating an axis of nonsymmorphic rotation~\cite{Kane12,Steinberg14}. 
A type-II Dirac semimetal fails to be an insulator due to the nonsymmorphic nature of its point group. 
The only continuously-tunable property this system has is the filling, since the Dirac points are pinned to TRIM. 
However, the standard filling anomaly of a metal of course vanishes in this case, since the filling is an integer. 
It is clear that, since a nonsymmorphic nature of the space group plays a crucial role, the corresponding anomaly 
term must involve the corresponding nonsymmorphic symmetry gauge field. 
We leave an in-depth exploration of this to future work, noting that some closely related ideas have been discussed in Refs.~\cite{Watanabe15,Watanabe16,Parameswaran19}.

\begin{acknowledgments}
CW, AH and XY acknowledge support from the Natural Sciences and Engineering Research Council (NSERC) of Canada.
AAB was supported by Center for Advancement of Topological Semimetals, an Energy Frontier Research Center funded by the U.S. Department of Energy Office of Science, Office of Basic Energy Sciences, through the Ames Laboratory under
contract DE-AC02-07CH11358. 
Research at Perimeter Institute is supported in part by the Government of Canada through the Department of Innovation, Science and Economic Development and by the Province of Ontario through the Ministry of Economic Development, Job Creation and Trade.
\end{acknowledgments}


\bibliography{references}

\end{document}